\documentclass[preprint2]{aastex}

\usepackage{amsfonts}
\usepackage{amsmath,amssymb}
\usepackage{graphicx}

\newcommand{\degree}{\ensuremath{^\circ}}

\begin{document}

\title{Analysis of large-scale anisotropy of ultra-high energy cosmic rays
in HiRes data}

\author{
R.~U.~Abbasi,\altaffilmark{1}
T.~Abu-Zayyad,\altaffilmark{1}
M.~Allen,\altaffilmark{1}
J.~F.~Amann,\altaffilmark{2}
G.~Archbold,\altaffilmark{1}
K.~Belov,\altaffilmark{1}
J.~W.~Belz,\altaffilmark{1}
D.~R.~Bergman,\altaffilmark{1,3}
S.~A.~Blake,\altaffilmark{1}
O.~A.~Brusova,\altaffilmark{1}
G.~W.~Burt,\altaffilmark{1}
C.~Cannon,\altaffilmark{1}
Z.~Cao,\altaffilmark{1,4}
W.~Deng,\altaffilmark{1}
Y.~Fedorova,\altaffilmark{1}
J.~Findlay,\altaffilmark{1}
C.~B.~Finley,\altaffilmark{5}
R.~C.~Gray,\altaffilmark{1}
W.~F.~Hanlon,\altaffilmark{1}
C.~M.~Hoffman,\altaffilmark{2}
M.~H.~Holzscheiter,\altaffilmark{2}
G.~Hughes,\altaffilmark{3}
P.~H\"{u}ntemeyer,\altaffilmark{1}
D.~Ivanov,\altaffilmark{3}
B.~F~Jones,\altaffilmark{1}
C.~C.~H.~Jui,\altaffilmark{1}
K.~Kim,\altaffilmark{1}
M.~A.~Kirn,\altaffilmark{6}
H.~Koers,\altaffilmark{7}
E.~C.~Loh,\altaffilmark{1}
M.~M.~Maestas,\altaffilmark{1}
N.~Manago,\altaffilmark{8}
L.~J.~Marek,\altaffilmark{2}
K.~Martens,\altaffilmark{1}
J.~A.~J.~Matthews,\altaffilmark{9}
J.~N.~Matthews,\altaffilmark{1}
S.~A.~Moore,\altaffilmark{1}
A.~O'Neill,\altaffilmark{5}
C.~A.~Painter,\altaffilmark{2}
L.~Perera,\altaffilmark{3}
K.~Reil,\altaffilmark{1}
R.~Riehle,\altaffilmark{1}
M.~D.~Roberts,\altaffilmark{9}
D.~Rodriguez,\altaffilmark{1}
M.~Sasaki,\altaffilmark{8}
S.~R.~Schnetzer,\altaffilmark{3}
L.~M.~Scott,\altaffilmark{3}
G.~Sinnis,\altaffilmark{2}
J.~D.~Smith,\altaffilmark{1}
P.~Sokolsky,\altaffilmark{1}
C.~Song,\altaffilmark{5}
R.~W.~Springer,\altaffilmark{1}
B.~T.~Stokes,\altaffilmark{1,3}
S.~R.~Stratton,\altaffilmark{3}
J.~R.~Thomas,\altaffilmark{1}
S.~B.~Thomas,\altaffilmark{1}
G.~B.~Thomson,\altaffilmark{1,3}
P.~Tinyakov,\altaffilmark{7}
D.~Tupa,\altaffilmark{2}
L.~R.~Wiencke,\altaffilmark{1}
A.~Zech\altaffilmark{3}
X.~Zhang\altaffilmark{5}\\
(The High Resolution Fly's Eye Collaboration)}


\altaffiltext{1}{University of Utah,
Department of Physics and High Energy Astrophysics Institute,
Salt Lake City, UT 84112, USA.}

\altaffiltext{2}{Los Alamos National Laboratory,
Los Alamos, NM 87545, USA.}

\altaffiltext{3}{Rutgers --- The State University of New Jersey,
Department of Physics and Astronomy, Piscataway, NJ 08854, USA.}

\altaffiltext{4}{Institute of High Energy Physics, Beijing, China}

\altaffiltext{5}{Columbia University, Department of Physics and
Nevis Laboratories, New York, NY 10027, USA.}

\altaffiltext{6}{Montana State University, Department of Physics,
Bozeman, MT, USA}

\altaffiltext{7}{Universit\'e Libre de Bruxelles (ULB), CP225, Bld. du
Triomphe, B-1050 Brussels, Belgium}

\altaffiltext{8}{University of Tokyo,
Institute for Cosmic Ray Research,
Kashiwa City, Chiba 277-8582, Japan.}

\altaffiltext{9}{University of New Mexico,
Department of Physics and Astronomy,
Albuquerque, NM 87131, USA.}

\begin{abstract}
Stereo data collected by the HiRes experiment over a six year period
are examined for large-scale anisotropy related to the inhomogeneous
distribution of matter in the nearby Universe. We consider the generic
case of small cosmic-ray deflections and a large number of sources
tracing the matter distribution. In this matter tracer model the
expected cosmic ray flux depends essentially on a single free
parameter, the typical deflection angle $\theta_s$. We find that the
HiRes data with threshold energies of 40 EeV and 57 EeV are incompatible
with the matter tracer model at a 95\% confidence level unless
$\theta_s>10^\circ$ and are compatible with an isotropic flux. The
data set above 10 EeV is compatible with both the matter tracer model
and an isotropic flux.
\end{abstract}



\maketitle

\section{Introduction}
The observation of the cutoff in the spectrum of Ultra-High Energy
Cosmic Rays (UHECRs) \citep{Abbasi:2007sv,Abraham:2008ru} as predicted
by \citet{Greisen:1966jv,Zatsepin:1966jv}  provides
compelling evidence for the shortening of the UHECR propagation length at
high energies. The highest energy events then must have come from
relatively close sources (within $250$ Mpc). At these length scales
the matter in the Universe is distributed inhomogeneously, being
organized into clusters and superclusters. One should, therefore, expect
the flux of highest-energy cosmic rays to be anisotropic.

In astrophysical scenarios, it is natural to assume that the number of
sources within $250$~Mpc is large, and that these sources trace the
distribution of matter. Under these assumptions, the anisotropy at
Earth depends only on the nature and size of UHECR
deflections. Measurement of the anisotropy, therefore, gives direct
experimental access to parameters that determine the deflections,
notably to the UHECR charge composition and cosmic magnetic fields.

Several investigations of anisotropy in arrival directions of UHECRs
have been previously undertaken.  At small angular scales,
correlations with different classes of putative sources were claimed
(e.g. \citealt{Gorbunov:2004bs,Abbasi:2005qy,Cronin:2007zz,Abraham:2007si}).
At larger angular scales and energies below 10 EeV possible anisotropy
towards the Galactic center was reported in
\citet{Hayashida:1998qb,Hayashida:1999ab,Bellido:2000tr}, but not
supported by more recent studies \citep{Santos:2007na}. At higher
energies, \citet{Stanev:1995my} found evidence against an isotropic
flux above 40 EeV through correlations with the supergalactic plane,
but this was not confirmed by other authors
\citep{Hayashida:1996bc,Kewley:1996zt, Bird:1998nu}. Finally, using
the Pierre Auger Observatory (PAO) data, \citet{Kashti:2008bw} have
found correlations between UHECR arrival directions and the
large-scale structure of the Universe which are incompatible with an
isotropic flux (see, however, \citealt{Koers:2008ba}).

In this paper, we analyze the data accumulated by the HiRes experiment
for anisotropy associated with the large-scale structure of the
Universe.  The HiRes experiment has been described previously
\citep{HiResStatus1999,Boyer:2002zz,Hanlon:2008}. It studied
ultrahigh energy cosmic rays from $10^{17.2}$ eV to $10^{20.2}$ eV
using the fluorescence technique.  HiRes operated two fluorescence
detectors located atop desert mountains in west-central Utah.  The
data set used in this study consists of events observed by both
detectors, analyzed jointly in what is commonly called ``stereo
mode''.  In this mode the angular resolution in cosmic rays' pointing
directions is about $0.8$ degrees, and the energy resolution is about
10\%.  The HiRes experiment operated in stereo mode between December,
1999, and April, 2006.  At the highest energies HiRes has the largest
data set in the Northern hemisphere. Large number of events, good
angular resolution (better than the bending angles expected from
Galactic and extragalactic magnetic fields) and the wide energy range
covered make the HiRes data particularly suitable for anisotropy
studies. The exact data set used in this study was described
previously in \citet{Abbasi:2008md}.

We consider here a generic model that assumes many sources within
$250$~Mpc tracing the distribution of matter, which we refer to as the
``matter tracer'' model. We also assume that deflections of UHECR do
not exceed the angular size of the nearby structures, that is
10-20$^\circ$. In this regime, both regular and random deflections in
magnetic fields can be modeled with a one-parameter distribution, for
which we take a Gaussian distribution centered at zero with width
$\theta_{\rm s}$. This width is treated as a free parameter, whose
value we aim to constrain from the data. Constraints on $\theta_{\rm
  s}$ may then be used to obtain information on the strength of
Galactic and extragalactic magnetic fields. In keeping with our
assumption of small deflections, we assume a proton composition in
this study, which is consistent with the $X_{\rm max}$ analysis based
on the same dataset (for confirmation see \citealt{Abbasi:2009nf}).

The HiRes data is compared to model predictions using the ``flux
sampling'' test put forward by \citet{Koers:2008ba}. This test has
good discrimination power at small statistics and is insensitive to
the details of deflections. The comparison is performed at three
different threshold energies that have been used in previous studies:
10 EeV, 40 EeV, and 57 EeV
\citep{Hayashida:1996bc,Abbasi:2005qy,Cronin:2007zz}.  An {\em a
  priori} significance of 5\%, corresponding to a confidence level
(CL) of 95\%, is chosen for this work.

The paper is organized as follows.  In section~\ref{section:modeling}
we discuss the modeling of UHECR arrival
directions. Section~\ref{section:data} concerns the HiRes data used in
the analysis, while section~\ref{section:fluxsampling} describes the
flux sampling method. We present our results in
section~\ref{sec:results} and conclude in
section~\ref{sec:conclusions}.

\section{Modeling of UHECR arrival directions}
\label{section:modeling}

\emph{Galaxy catalog ---} The distribution of matter in the local
Universe is modeled with the 2 Micron All-Sky Redshift Survey (2MRS;
\citealt{2MRS}) galaxy catalog, using galaxies as samplers of the
underlying matter density field.\footnote{This sample was kindly
  provided by John Huchra.}  The 2MRS is a flux-limited sample of
galaxies, that is, the sample containing all galaxies with
observed magnitude in the $K_s$ band $m \leq 11.25$. It contains
spectroscopically measured redshifts for all but a few galaxies.  A
number of cuts have been applied to the galaxy sample.  First, the
Galactic plane, where the sample is incomplete, has been excluded from
the sample by removing all galaxies with $|b|<10\degree$.  Second,
objects with $D<5$ Mpc are removed because such objects should be
treated on an individual basis.\footnote{This corresponds to the
  \emph{ad hoc} assumption that there are no UHECR sources within 5
  Mpc.  Different analyses are more appropriate to test this
  possibility.}  Finally, the catalog is cut at 250~Mpc because the
sample becomes too sparse. The resulting sample provides an accurate
statistical description at smearing angles $\theta_s>2^\circ$.  The
flux from sources beyond 250~Mpc is taken to be isotropic.  A total of
15508 galaxies remain in the HiRes field of view after the cuts.  To
compensate for observational selection effects in the (flux-limited)
2MRS catalog, weights $ w^{\rm cat}_i$ are assigned to the galaxies
with the sliding-box method as described in \citet{Koers:2009pd}.

\begin{figure}
\includegraphics[angle=-90, width=8cm]{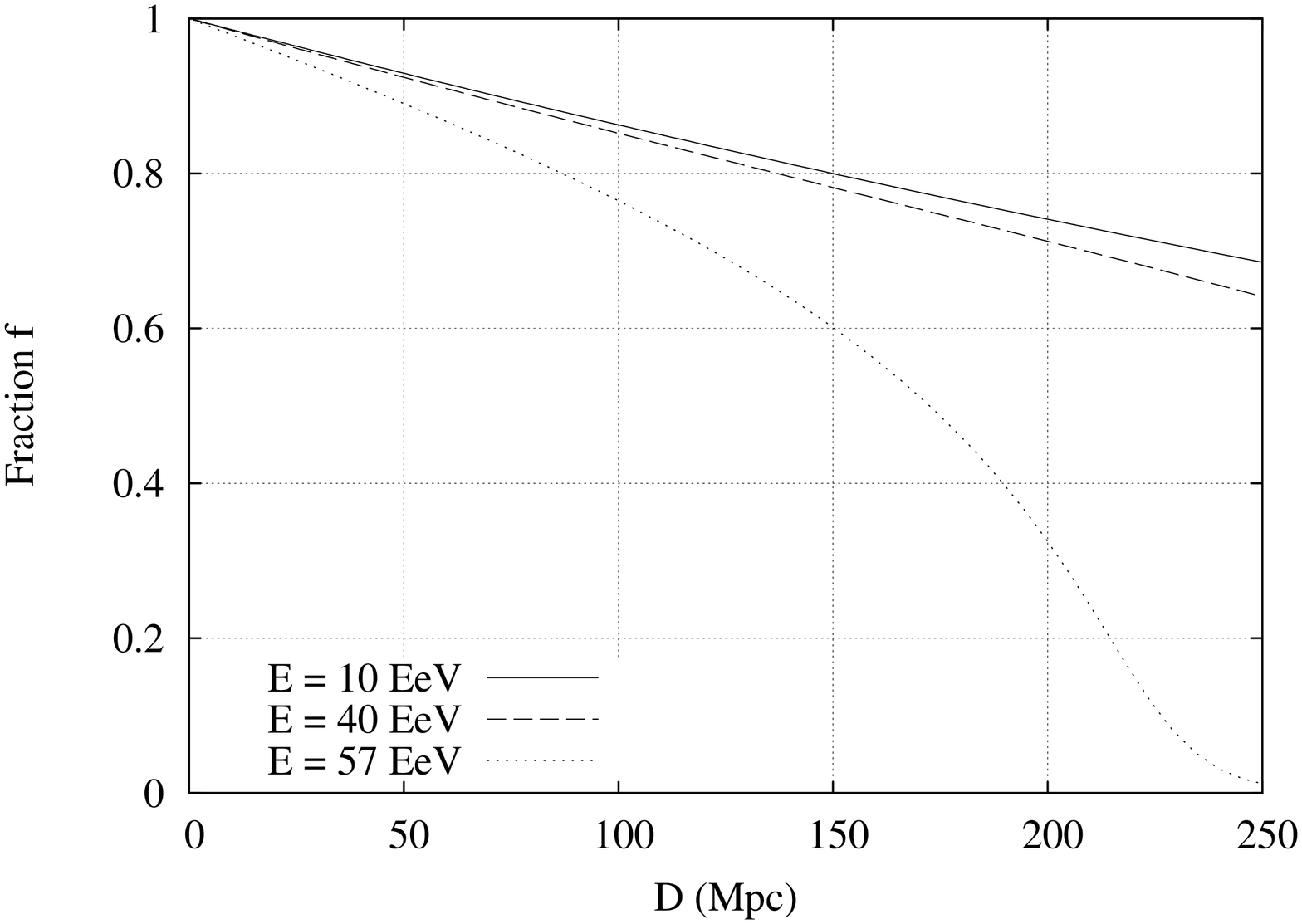}
\includegraphics[angle=-90, width=8cm]{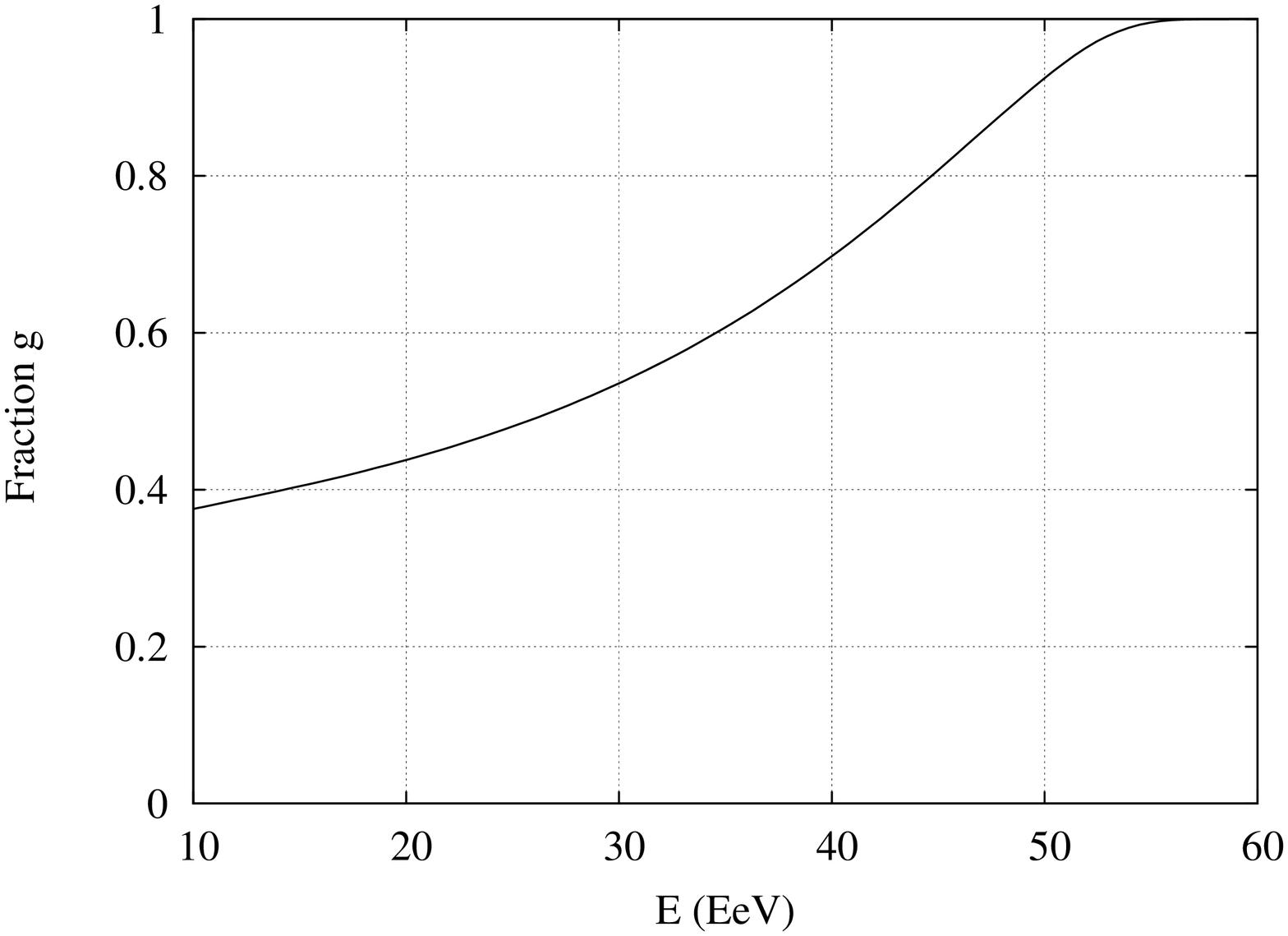}
\caption{\label{fig:ffunc} {\em Top panel:} Fraction $f$ of integral
  CR flux that survives after interactions with the CMB and
  cosmological redshift as a function of distance $D$ for threshold
  energies 10 EeV (solid line), 40 EeV (dashed), and 57 EeV (dotted).
  {\em Bottom panel:} Fraction $g$ of total flux that is produced by
  sources within $250$~Mpc, as a function of energy $E$.  }
\end{figure}

\emph{Energy loss ---} UHECR fluxes are affected by energy loss due to
redshift and interactions with the Cosmic Microwave Background (CMB).
To account for the resulting flux suppression, the integral flux,
$\varphi_i$, from a single source is expressed as follows: 
\begin{equation}
\label{eq:ptflux}
\varphi_i (E, D_i)  = \frac{J (E) f (E, D_i)}{4 \pi D_i^2} \, ,
\end{equation}
where $E$ is the threshold energy, $D_i$ is the source distance, $J$
stands for the integral injection spectrum at the source, and $f$
represents the flux fraction that remains after interactions and
redshift.  We take an injection spectrum $J(E) \propto E^{-1.2}$
extending to very high energies. The function $f$ is determined using
a numerical propagation code based on the continuous loss
approximation that is described in \citet{Koers:2008hv,Koers:2008ba}.
Energy loss due to interactions with the extragalactic background
light is neglected.  In Figure~\ref{fig:ffunc}, top panel, the
fraction $f$ is shown as a function of distance for the different
energies considered in this work.

The strength of the isotropic flux component that is added to account
for sources beyond 250 Mpc also depends on UHECR energy loss.  Using
the computer code described in the previous paragraph, we estimate
the fraction $g$ of total flux contributed by sources within 250 Mpc
to be 0.4, 0.7, and 1.0 for threshold energies $E=10$ EeV, 40 EeV, and
57 EeV, respectively (see Figure 1, bottom panel).

\emph{Deflections ---} UHECR protons (as well as nuclei) are deflected
by Galactic and intergalactic magnetic fields. These deflections are
taken into account by an angular smearing procedure, which replaces
the point-source flux, $\varphi$, by a flux density distribution:
\begin{equation}
\varphi_i \to \varphi_i \, w^s (\theta_i) \, ,
\end{equation}
where $w^s (\theta_i)$ represents the probability density that an
UHECR is deflected by $\theta_i$, the angle between galaxy $i$ and the
line of sight.  This procedure also accounts for the detector's
angular resolution and prevents unphysical fluctuations due to the use
of a catalog of point sources.  In the absence of detailed knowledge
on the structure of Galactic and extragalactic magnetic fields, we
adopt a simple Gaussian probability density distribution with
characteristic angle, $\theta_{\rm s}$. This angle is treated as a free
model parameter.  The Gaussian distribution is a fair approximation
when the deflections are small. For large deflections, details on the
structure of the Galactic and extragalactic magnetic fields become
important. Accounting for these details goes beyond the scope of the
present study.

\emph{Exposure ---} The HiRes exposure is modeled using our Monte
Carlo simulation of the experiment \citep{Abbasi:2006mb,
  Bergman:2006vt}.  The aim of this simulation was to create a set of
Monte Carlo events that would be in all essences identical to the
actual data.  In making the simulation we put in the properties of
cosmic ray air showers as measured by previous experiments
\citep{Bird:1993yi, AbuZayyad:2000zz, AbuZayyad:2000ay,
  Abbasi:2004nz}.  We used cosmic ray showers generated by the Corsika
and QGSJet programs \citep{Heck:1998vt, Kalmykov:1997te} and simulated
the generation of fluorescence light (see references in
\citealt{Abbasi:2007sv}) and its propagation through the atmosphere
(see references in \citealt{Abbasi:2007sv}).  A complete simulation of
the optics and electronics (trigger and front-end electronics) of our
detectors was performed.  The result was an excellent simulation of
our experiment as evidenced by the very good agreement between data
and simulated events in the distribution of all kinematic variables,
e.g. zenith angle, impact parameter to detector, etc. By assigning
Monte Carlo events times of occurrence taken from the actual on-time
of the experiment we are able to calculate the exposure on the sky
very accurately.

\emph{Model flux maps ---} The integral UHECR flux from a given
direction is expressed as follows:
\begin{equation}
\Phi  =  \sum_i  \varphi_i  \, w^{\rm cat}_i  \, w^{\rm s} (\theta_i) 
+ \Phi_{\rm iso} \, ,
\end{equation}
where $i$ enumerates galaxies in the 2MRS sample, $w^{\rm cat}_i$
denotes the weight assigned to galaxy $i$ in the catalog, $w^{\rm s}
(\theta_i)$ is the deflection probability distribution, and
$\Phi_{\rm iso}$ is the UHECR flux arising from sources beyond 250
Mpc.

\begin{figure}
\includegraphics[width=8cm]{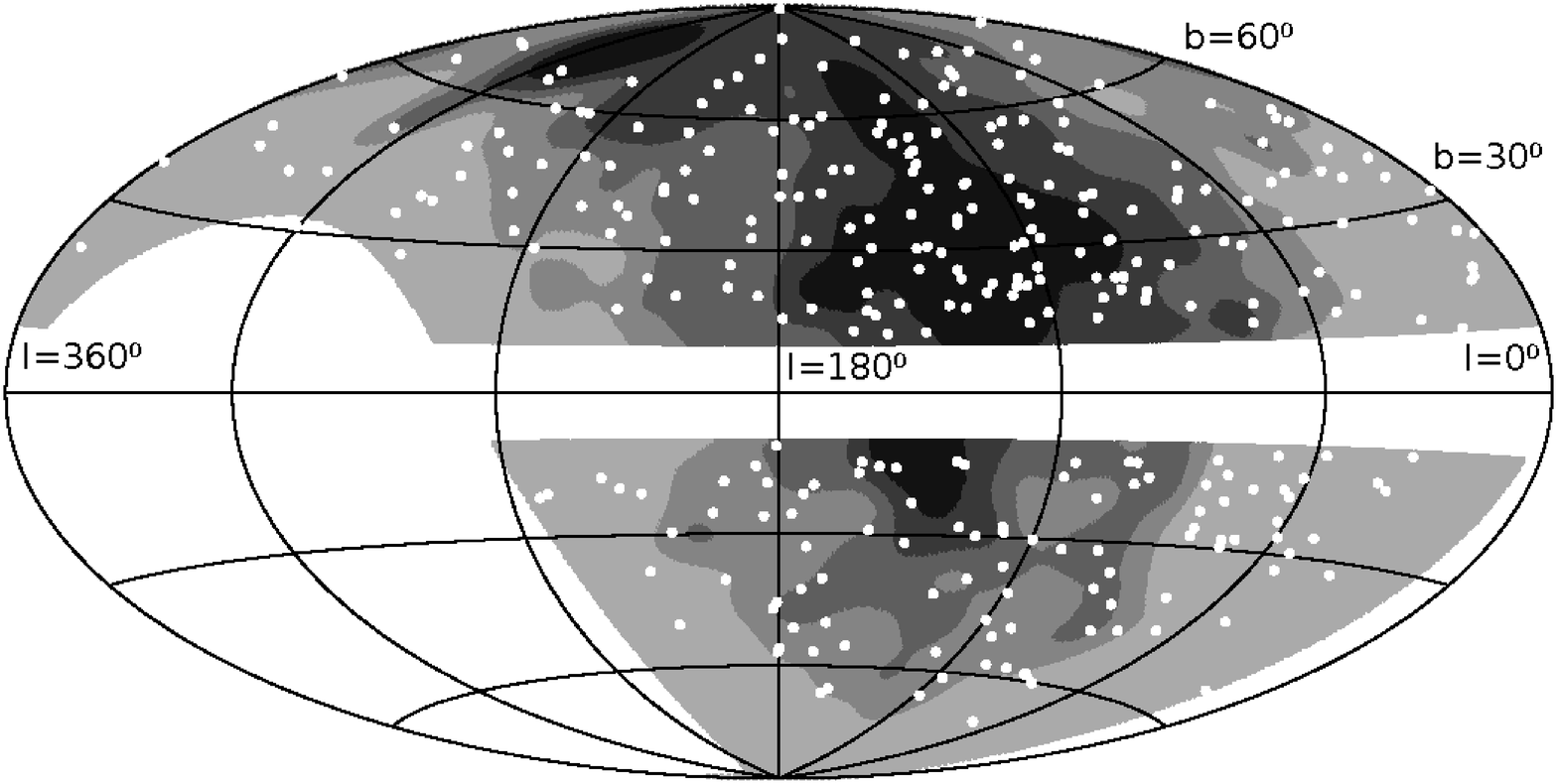}
\vspace{0.2cm}

\includegraphics[width=8cm]{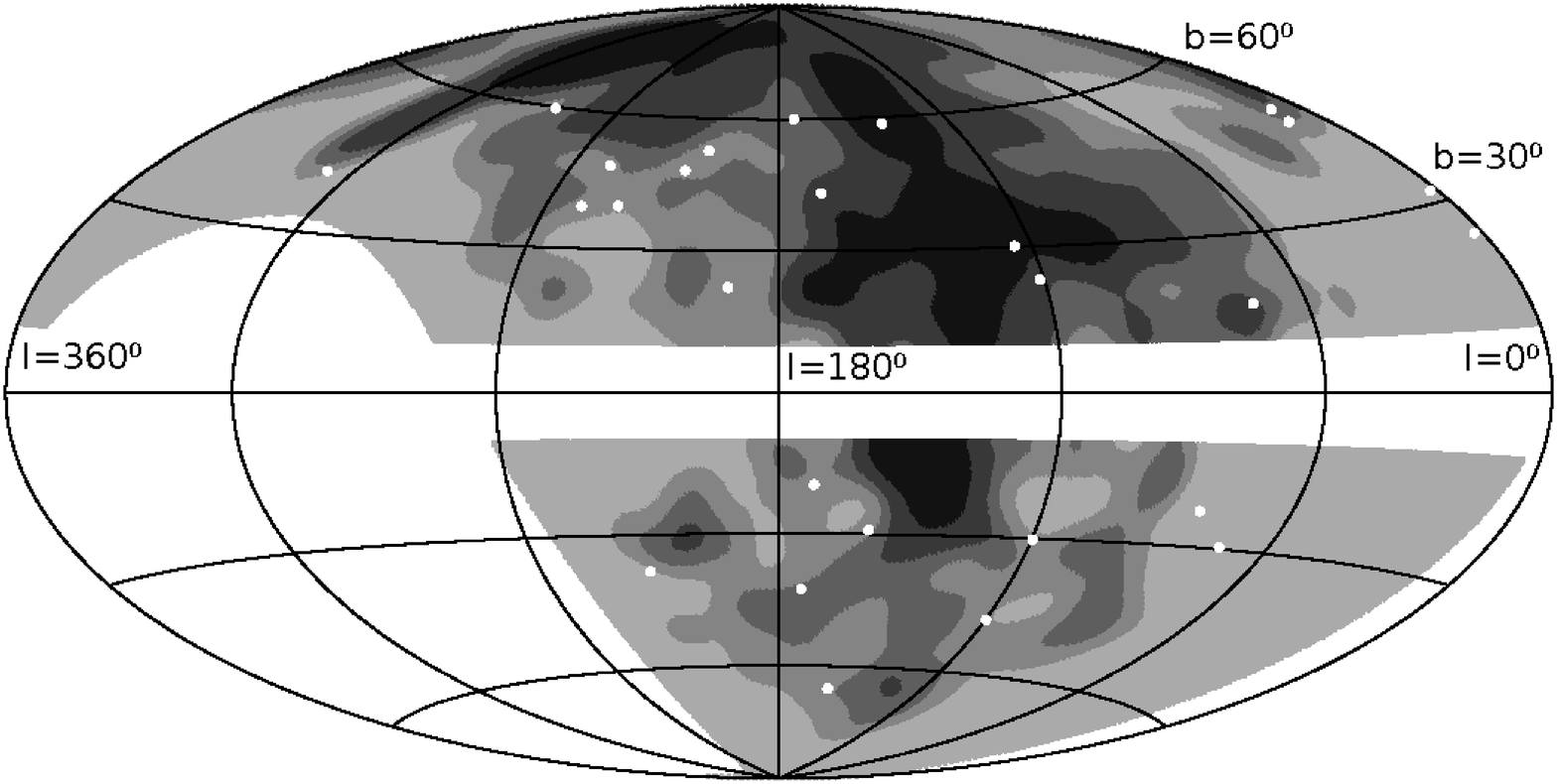}
\vspace{0.2cm}

\includegraphics[width=8cm]{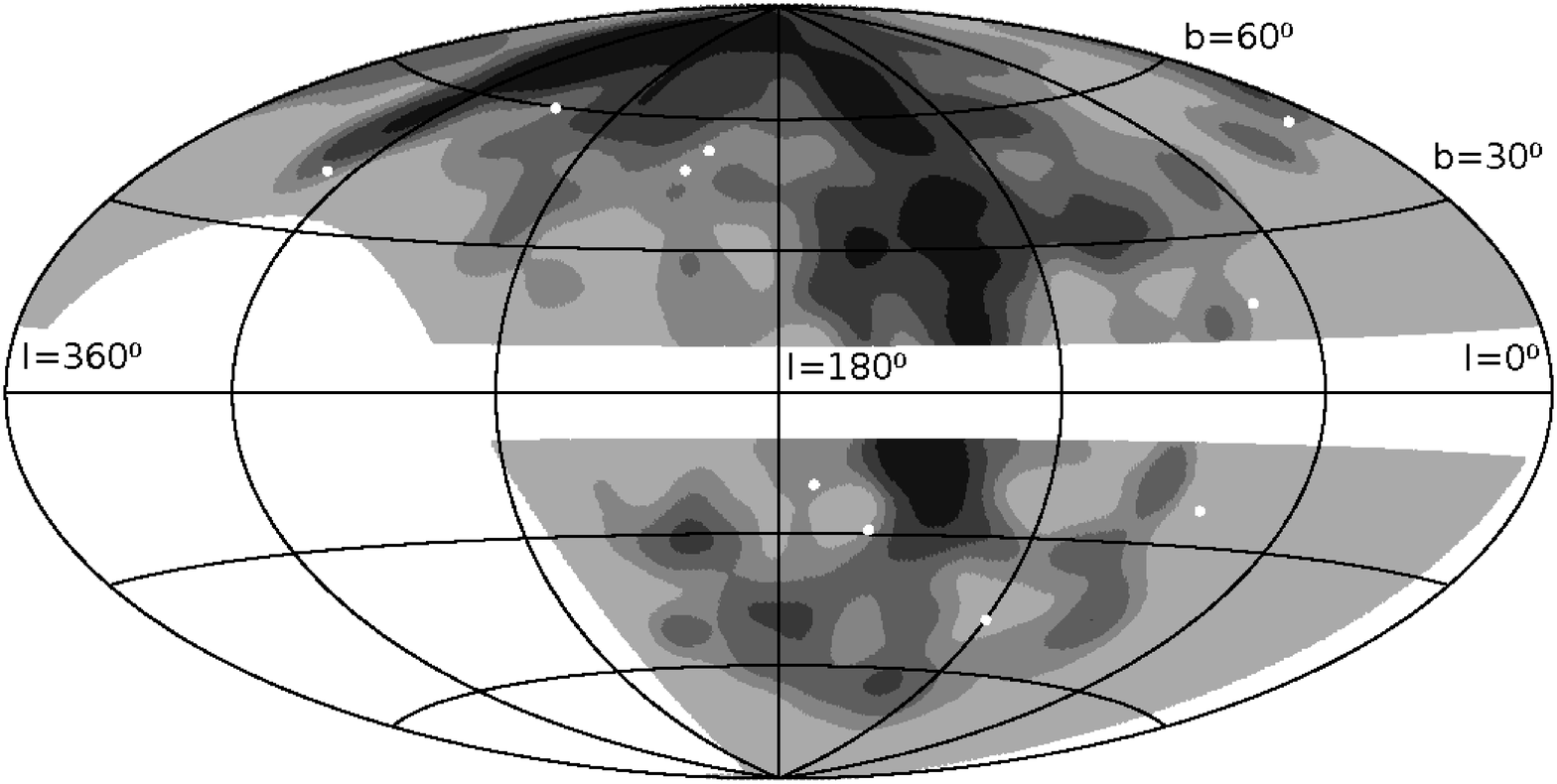}
\vspace{0.2cm}

\caption{\label{fig:skymap:struct} Hammer projection (galactic
  coordinates) of $\widetilde{\Phi}$ (flux times exposure) with
  threshold energies 10 EeV (top panel), 40 EeV (middle), and 57 EeV
  (bottom).  Darker gray indicates a higher value; the bands are
  chosen such that each band contains 1/5 of the total flux (weighted
  with exposure).  Excluded regions, viz. the galactic plane
  ($|b|<10\degree$) and the region outside the HiRes~ field of view,
  are shown in white. White dots indicate HiRes events.  All maps are
  produced with $\theta_{\rm s} = 6\degree$.}
\end{figure}

The probability to observe a CR from a given direction is proportional
to the product of flux and exposure. We denote this probability as
\begin{equation}
\widetilde{\Phi} = \Phi \, \Xi \, , 
\end{equation}
where $\Xi$ stands for exposure. In Figure \ref{fig:skymap:struct}, the
distribution of $\widetilde{\Phi}$ over the sky is shown for three
different threshold energies. The contrast in the flux distributions
becomes more pronounced with increasing energy. Also shown are the
arrival directions of UHECRs in the HiRes data to which the model flux
has to be compared.

\section{Data}
\label{section:data}

The data set used in this study was described previously in
\citet{Abbasi:2008md}, including selection criteria and a correction
to the energy scale.  Our sample of the 2MRS catalog does not cover
the region near the Galactic plane with $|b|<10\degree$. We therefore
removed cosmic ray events with $|b|<10\degree$ from the analysis. The
resulting sample contains: 
\begin{itemize}
\item 309 events with $E>10$~EeV;
\item 27 events with $E>40$~EeV;
\item 10 events with $E>57$~EeV.
\end{itemize}
The arrival directions of these events are shown as white dots in
Fig.~\ref{fig:skymap:struct}.

\section{Statistical test}
\label{section:fluxsampling}

The compatibility of a model flux map with the set of UHECR arrival
directions is quantified by the flux sampling method introduced by
\citet{Koers:2008ba}. The idea of the method is as follows. To any set
of arrival directions one associates the set of flux values that are
obtained by sampling a given flux map (such as the map shown in
fig.~\ref{fig:skymap:struct}),
i.e. by extracting the flux values at corresponding points on the
sphere. The two-dimensional distribution of arrival directions thus
translates into a one-dimensional distribution of flux values. If the
reference model is true, this flux distribution will tend to high
values since events fall preferentially into regions where the model
flux (times exposure) is high. If, on the other hand, the reference
model is not true, the flux distribution is more uniform because the
correlation between arrival directions and regions of high model flux
is (partly) destroyed. By comparing the flux distribution to a model
flux distribution, the compatibility between a set of arrival
directions and model predictions can be quantified. This comparison is
performed by the Kolmogorov-Smirnov (KS) test, which yields a test
statistic $D$. The relevant statistical quantities, in particular powers
and $p$-values, are computed from the distribution of this test
statistic. Note that this test does not involve any additional
parameters like bin size.

\begin{figure}
\includegraphics[angle=-90, width=8cm]{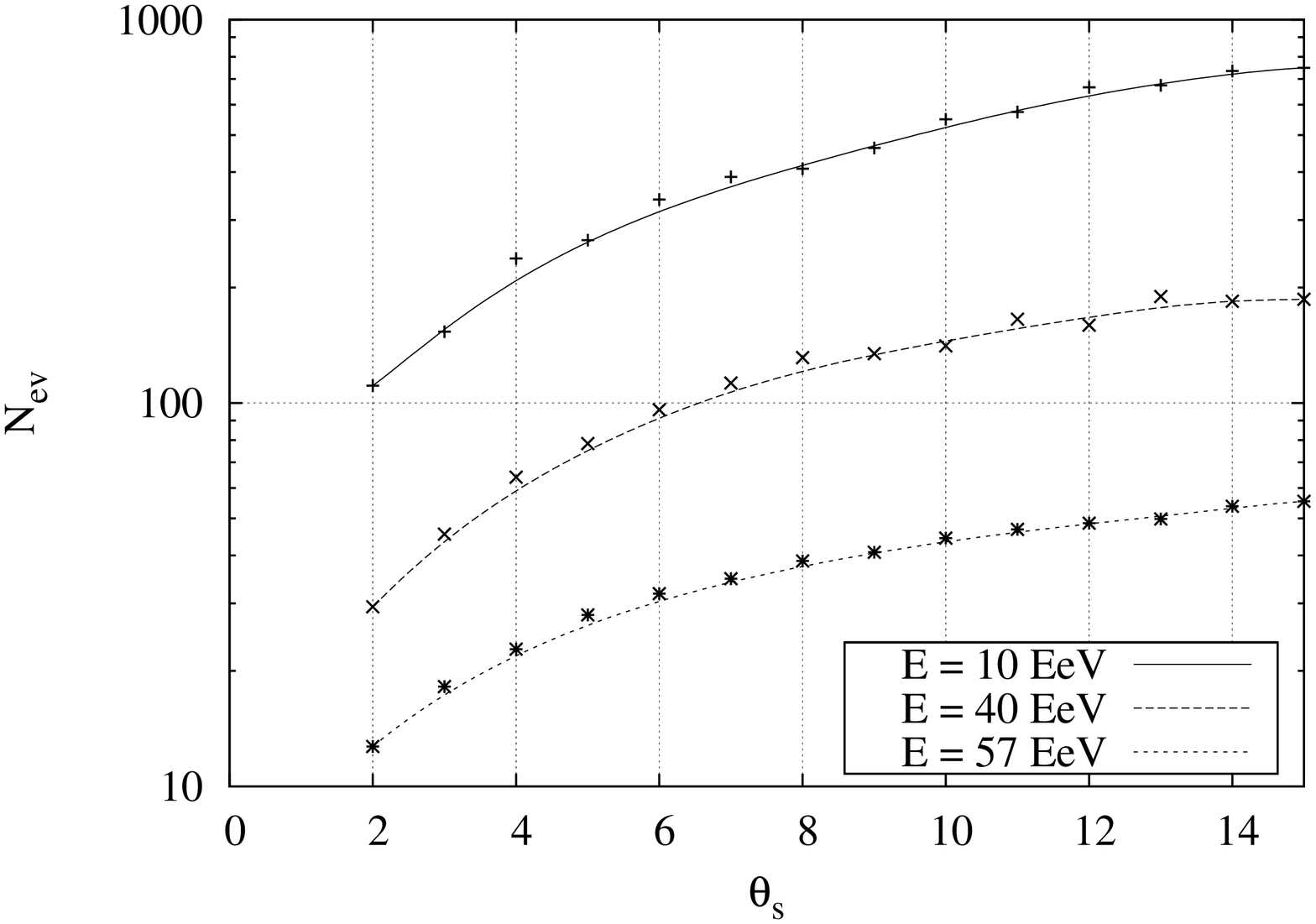}
\caption{\label{fig:power-Nvsths}
Number of events required for a 50\% probability to rule
out, at 95\% CL, the matter tracer model if the true flux
is isotropic.}
\end{figure}

The ability of the test to discriminate between models may be
quantified in terms of the statistical power, $P$, i.e. the probability
to rule out, at a given confidence level, the reference model when an
alternative model is true. Within numerical uncertainties, the
statistical power is equal to the fraction of event sets generated
under the alternative model that leads to rejecting the reference
model.  Figure~\ref{fig:power-Nvsths} shows the number of events
required for a power $P=0.5$ (i.e., a 50\% probability) to rule out
(at 95\% CL) the matter tracer model when the true flux is
isotropic. The number of events increases with increasing smearing
angle and decreasing energy: the decreasing flux contrasts in the
matter tracer model call for an increase in statistics to achieve the
same discriminatory power. Observe that the event numbers indicated in
Figure~\ref{fig:power-Nvsths} are of the same order as the data
analyzed in this work.  We thus expect that there is sufficient data
to obtain meaningful constraints at 95\% CL. 

\section{Results}
\label{sec:results}

\subsection{Scan over smearing angles}

\begin{figure}
\includegraphics[angle=-90, width=8cm]{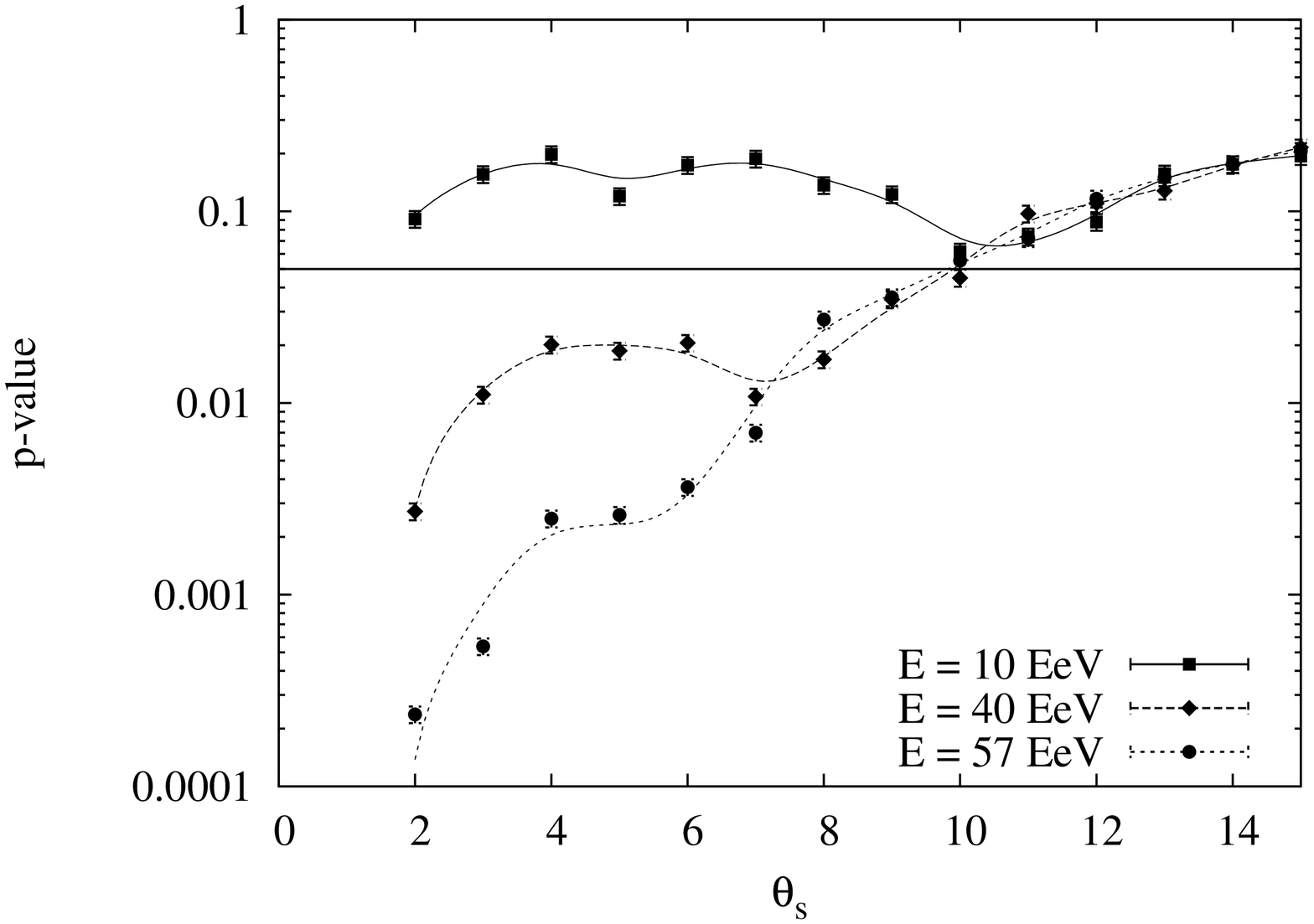}
\includegraphics[angle=-90, width=8cm]{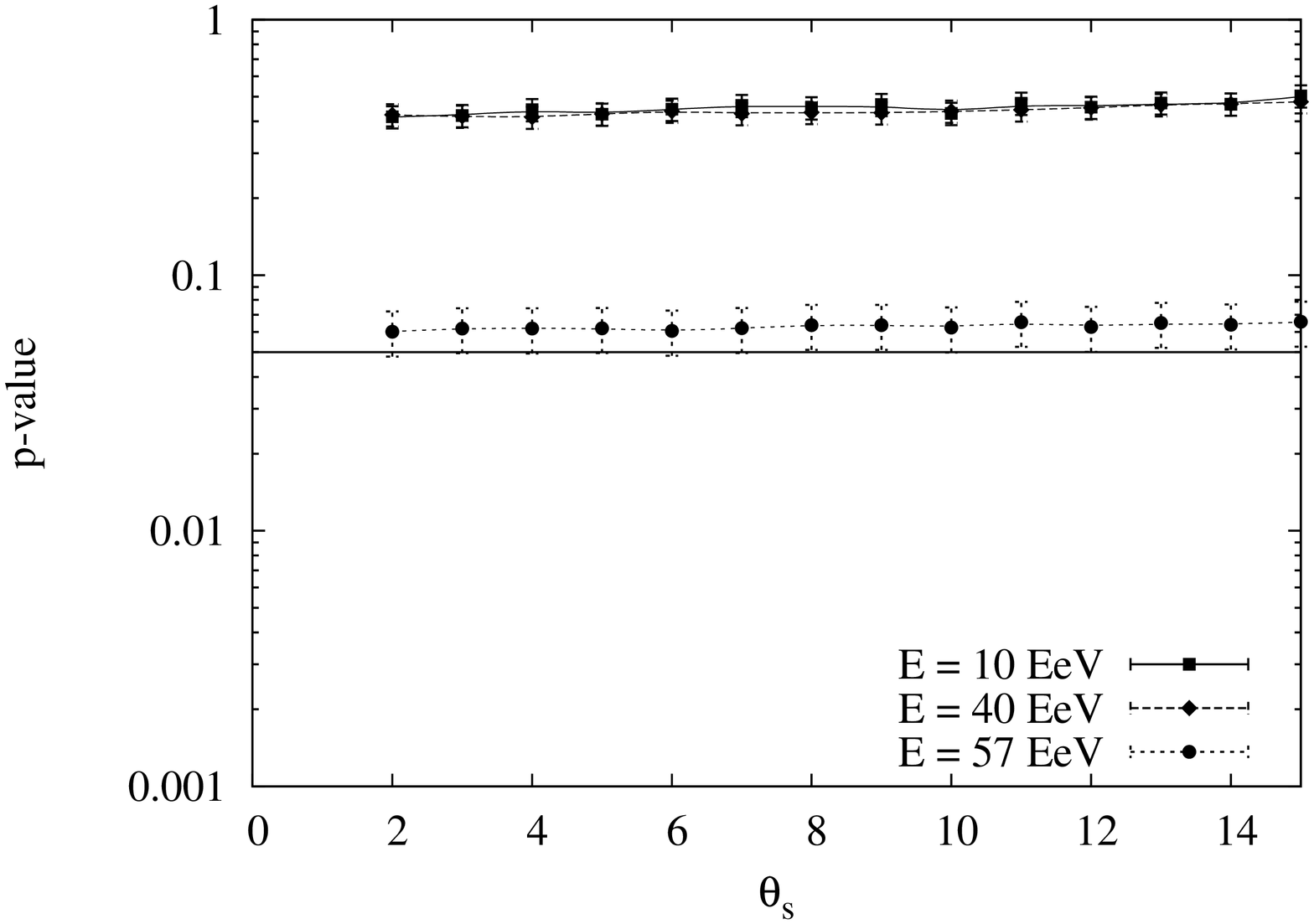}

\caption{\label{fig:pvalues} The dependence of $p$-value indicating
  the level of (in)compatibility between HiRes data and model
  predictions on the smearing angle $\theta_{\rm s}$.  Solid lines
  indicate a $p$-value equal to 0.05, below which the model is ruled
  out at 95\% CL. The points represent numerical results (with
  estimated uncertainties of 20\%); the lines are smooth
  interpolations between these points. Top panel: data vs. matter-tracer model;
  bottom panel: data vs. isotropic distribution. 
}
\end{figure}

The level of compatibility between data and model predictions
is quantified by a $p$-value, which represents the model probability
of obtaining a measurement that is at least as extreme as
the actual measurement. With our {\em a priori} choice of
significance, a $p$-value smaller than $0.05$ rules out the model. 
The probability of falsely excluding
the model is then $5$\%, translating to a CL of $95$\%.

Figure \ref{fig:pvalues} shows the $p$-values obtained by the flux
sampling method for the HiRes data and predictions of the matter
tracer model.  The smearing angle, $\theta_{\rm s}$, is treated as a
free parameter. That is, at each value of $\theta_{\rm s}$ and each
threshold energy a flux map is generated and compared to the HiRes
data as described above. The results can be summarized as follows:
\begin{itemize}
\item[\emph{(a)}] For the threshold energies of 40 EeV and 57 EeV, the
  tests show  disagreement between data and the matter
  tracer model for $\theta_{\rm s} \leq 10\degree$. Within this
  parameter range, a source distribution tracing the distribution of
  matter is excluded at a 95\% confidence level.
\item[\emph{(b)}] For the threshold energy of 10 EeV, the test shows
  agreement between data and the matter tracer model.
\end{itemize}
The incompatibility between data and matter tracer model is
illustrated by the non-correlation between the observed arrival
directions and regions of high model flux shown in the two lower
panels of Figure \ref{fig:skymap:struct}.

We have also tested the data for compatibility with an isotropic model
flux and found no disagreement, at 95\% CL, for any of the three tested
threshold energies (the data with threshold energy 57 EeV are
marginally consistent with an isotropic flux).

\subsection{Case study: $E = 57$ EeV, $\theta_{\rm s} = 3.2 \degree$}
At energy threshold $E > 57$~EeV a correlation between the arrival
directions of UHECRs and the location of AGNs contained in the
12$^{\rm th}$ edition of the V{\'e}ron-Cetty \& V{\'e}ron catalog
\citep{2006A&A...455..773V} was reported by the PAO
\citep{Cronin:2007zz, Abraham:2007si}. This correlation was found to
be maximal for $\psi= 3.2\degree$, where $\psi$ denotes the maximum
angular distance between UHECRs and AGNs.  In the Northern hemisphere,
correlation with AGN was not confirmed by the HiRes experiment
\citep{Abbasi:2008md}.

Since AGNs are tracers of the distribution of matter in the Universe,
the PAO result is suggestive of a more general correlation between
UHECRs and the local structure of the Universe on an angular scale of
a few degrees. The methods presented in this paper allow a check on
the existence of such correlations in the HiRes data.

\begin{figure}
\includegraphics[angle=-90, width=8cm]{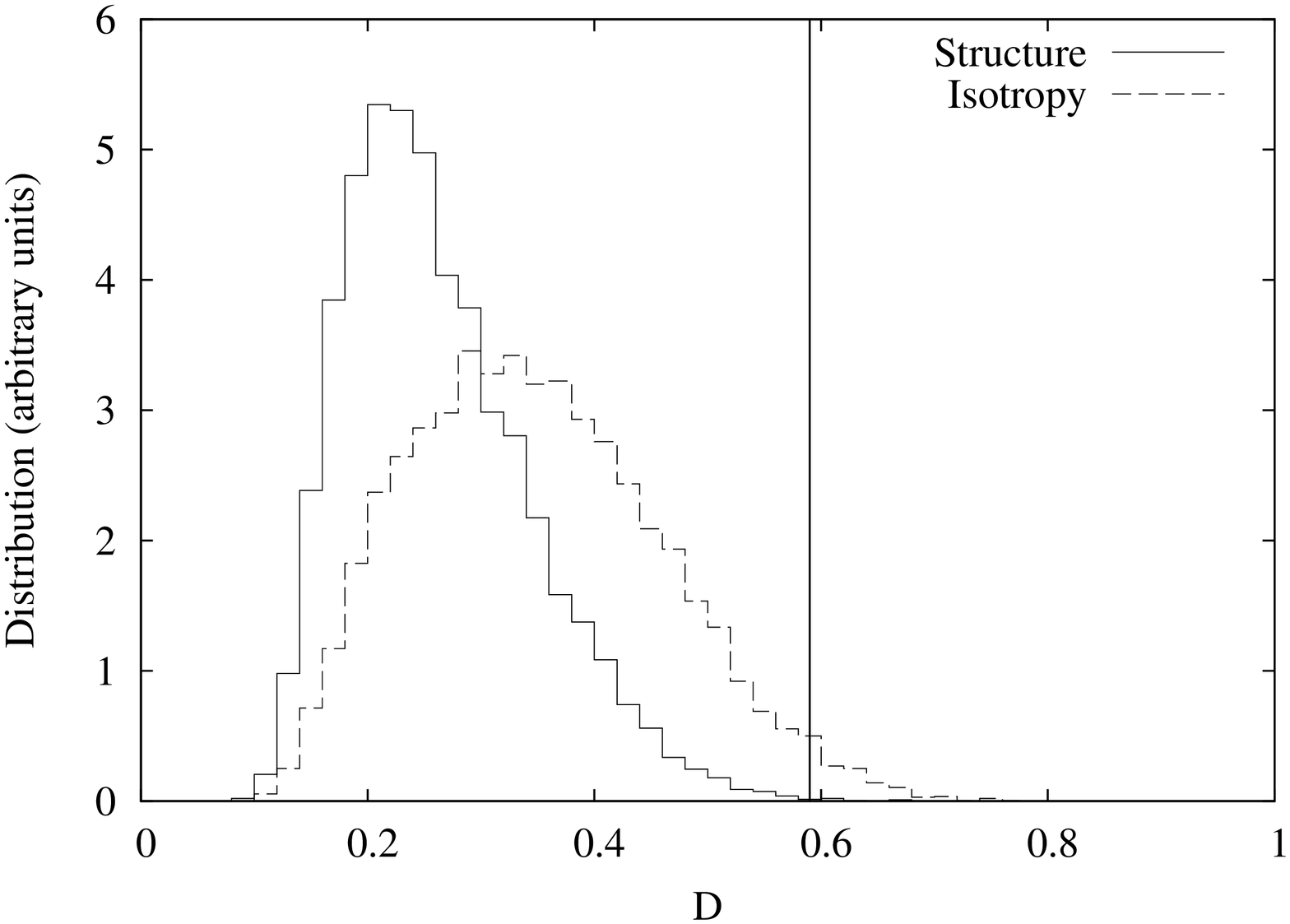}
\caption{\label{fig:Ddist} Model distribution of test statistic $D$
  when testing the matter tracer model, for both the matter tracer
  model (``Structure'') and an isotropic flux distribution
  (``Isotropy''). Here $E=57$ EeV and $\theta_{\rm s} =3.2\degree$.
  The vertical line indicates the observed value $D_{\rm obs}=0.59$.}
\end{figure}

The results presented in the previous section disfavor a correlation
between UHECRs and the local structure of the Universe. In fact, the
flux sampling test yields $p$-values smaller than $10^{-2}$ for the
matter tracer model with $\theta_{\rm s} \lesssim 6\degree$, with a
$p$-value of $7\times 10^{-4}$ for $\theta_{\rm s} = 3.2\degree$.
(Note that $\theta_{\rm s}$ is not in $1:1$ correspondence with
$\psi$; both quantities are however representative of the angular
scale of the problem).  Focusing on the case of $\theta_{\rm s} =
3.2\degree$ in more detail, Figure \ref{fig:Ddist} shows the
distribution of the test statistic $D$ for the matter tracer model and
for an isotropic flux for this smearing angle. The vertical line shows
the value $D_{\rm obs} = 0.59$ obtained for the HiRes data. This
demonstrates the strong incompatiblity between HiRes data and the
matter tracer model for smearing angle $\theta_{\rm s} = 3.2\degree$
and threshold energy $E=57$ EeV.

\section{Conclusions}
\label{sec:conclusions}

To summarize, we have confronted the stereo data collected by
the HiRes experiment with predictions of the matter tracer model, a
generic model of cosmic ray origin and propagation. The model assumes
a large number of cosmic ray sources within $250$~Mpc whose
distribution traces that of matter, and relatively small deflections
characterized by a single parameter, the typical deflection angle
$\theta_s$. We have found that the HiRes data with energy thresholds
$E=40$~EeV and $E=57$~EeV are incompatible with the matter tracer
model for $\theta_s<10^\circ$ at 95\%~CL.  With an energy threshold
$E=10$~EeV the HiRes data are compatible with the matter tracer model.
At all three energy thresholds, the data are compatible with an
isotropic flux at 95\%~CL.

In the present analysis we have treated the deflections as random and
Gaussian, which is only appropriate for small deflection angles and
limited number of events. The actual deflections are expected to
contain a coherent component due to the Galactic magnetic field. With
the accumulation of UHECR events by PAO in the Southern hemisphere and
by Telescope Array in the Northern hemisphere, our analysis
will become sensitive to the nature of deflections and thus, with
proper modifications of the statistical procedure, may give direct
access to the parameters of cosmic magnetic fields.

\section*{Acknowledgments}

This work is supported by the National Science Foundation under
contracts NSF-PHY-9321949, NSF-PHY-9322298, NSF-PHY-9974537,
NSF-PHY-0071069, NSF-PHY-0098826, NSF-PHY-0140688, NSF-PHY-0245328,
NSF-PHY-0307098, and NSF-PHY-0305516, by Department of Energy grant
FG03-92ER40732, by the BSP under IUAP VI/11, by the FNRS contract
1.5.335.08 and by the IISN contract 4.4509.10. We gratefully
acknowledge the contribution from the technical staffs of our home
institutions and thank the University of Utah Center for High
Performance Computing for their contributions. The cooperation of
Colonels E. Fisher, G. Harter, and G. Olsen, the US Army and the
Dugway Proving Ground staff is appreciated.

\bibliographystyle{apsrev}

\begin{thebibliography}{35}
\expandafter\ifx\csname natexlab\endcsname\relax\def\natexlab#1{#1}\fi
\expandafter\ifx\csname bibnamefont\endcsname\relax
  \def\bibnamefont#1{#1}\fi
\expandafter\ifx\csname bibfnamefont\endcsname\relax
  \def\bibfnamefont#1{#1}\fi
\expandafter\ifx\csname citenamefont\endcsname\relax
  \def\citenamefont#1{#1}\fi
\expandafter\ifx\csname url\endcsname\relax
  \def\url#1{\texttt{#1}}\fi
\expandafter\ifx\csname urlprefix\endcsname\relax\def\urlprefix{URL }\fi
\providecommand{\bibinfo}[2]{#2}
\providecommand{\eprint}[2][]{\url{#2}}

\bibitem[{\citenamefont{Abbasi et~al.}(2008{\natexlab{a}})}]{Abbasi:2007sv}
\bibinfo{author}{\bibfnamefont{R.}~\bibnamefont{Abbasi}} \bibnamefont{et~al.}
  (\bibinfo{collaboration}{HiRes}), \bibinfo{journal}{Phys. Rev. Lett.}
  \textbf{\bibinfo{volume}{100}}, \bibinfo{pages}{101101}
  (\bibinfo{year}{2008}{\natexlab{a}}), \eprint{astro-ph/0703099}.

\bibitem[{\citenamefont{Abraham et~al.}(2008{\natexlab{a}})}]{Abraham:2008ru}
\bibinfo{author}{\bibfnamefont{J.}~\bibnamefont{Abraham}} \bibnamefont{et~al.}
  (\bibinfo{collaboration}{Pierre Auger}), \bibinfo{journal}{Phys. Rev. Lett.}
  \textbf{\bibinfo{volume}{101}}, \bibinfo{pages}{061101}
  (\bibinfo{year}{2008}{\natexlab{a}}), \eprint{0806.4302}.

\bibitem[{\citenamefont{Greisen}(1966)}]{Greisen:1966jv}
\bibinfo{author}{\bibfnamefont{K.}~\bibnamefont{Greisen}},
  \bibinfo{journal}{Phys. Rev. Lett.} \textbf{\bibinfo{volume}{16}},
  \bibinfo{pages}{748} (\bibinfo{year}{1966}).

\bibitem[{\citenamefont{Zatsepin and Kuzmin}(1966)}]{Zatsepin:1966jv}
\bibinfo{author}{\bibfnamefont{G.~T.} \bibnamefont{Zatsepin}} \bibnamefont{and}
  \bibinfo{author}{\bibfnamefont{V.~A.} \bibnamefont{Kuzmin}},
  \bibinfo{journal}{JETP Lett.} \textbf{\bibinfo{volume}{4}},
  \bibinfo{pages}{78} (\bibinfo{year}{1966}).

\bibitem[{\citenamefont{Gorbunov et~al.}(2004)\citenamefont{Gorbunov, Tinyakov,
  Tkachev, and Troitsky}}]{Gorbunov:2004bs}
\bibinfo{author}{\bibfnamefont{D.~S.} \bibnamefont{Gorbunov}},
  \bibinfo{author}{\bibfnamefont{P.~G.} \bibnamefont{Tinyakov}},
  \bibinfo{author}{\bibfnamefont{I.~I.} \bibnamefont{Tkachev}},
  \bibnamefont{and} \bibinfo{author}{\bibfnamefont{S.~V.}
  \bibnamefont{Troitsky}}, \bibinfo{journal}{JETP Lett.}
  \textbf{\bibinfo{volume}{80}}, \bibinfo{pages}{145} (\bibinfo{year}{2004}),
  \eprint{astro-ph/0406654}.

\bibitem[{\citenamefont{Abbasi et~al.}(2006)}]{Abbasi:2005qy}
\bibinfo{author}{\bibfnamefont{R.~U.} \bibnamefont{Abbasi}}
  \bibnamefont{et~al.} (\bibinfo{collaboration}{HiRes}),
  \bibinfo{journal}{Astrophys. J.} \textbf{\bibinfo{volume}{636}},
  \bibinfo{pages}{680} (\bibinfo{year}{2006}), \eprint{astro-ph/0507120}.

\bibitem[{\citenamefont{Abraham et~al.}(2007)}]{Cronin:2007zz}
\bibinfo{author}{\bibfnamefont{J.}~\bibnamefont{Abraham}} \bibnamefont{et~al.}
  (\bibinfo{collaboration}{Pierre Auger}), \bibinfo{journal}{Science}
  \textbf{\bibinfo{volume}{318}}, \bibinfo{pages}{938} (\bibinfo{year}{2007}),
  \eprint{0711.2256}.

\bibitem[{\citenamefont{Abraham et~al.}(2008{\natexlab{b}})}]{Abraham:2007si}
\bibinfo{author}{\bibfnamefont{J.}~\bibnamefont{Abraham}} \bibnamefont{et~al.}
  (\bibinfo{collaboration}{Pierre Auger}), \bibinfo{journal}{Astropart. Phys.}
  \textbf{\bibinfo{volume}{29}}, \bibinfo{pages}{188}
  (\bibinfo{year}{2008}{\natexlab{b}}), \eprint{0712.2843}.

\bibitem[{\citenamefont{Hayashida
  et~al.}(1999{\natexlab{a}})}]{Hayashida:1998qb}
\bibinfo{author}{\bibfnamefont{N.}~\bibnamefont{Hayashida}}
  \bibnamefont{et~al.} (\bibinfo{collaboration}{AGASA}),
  \bibinfo{journal}{Astropart. Phys.} \textbf{\bibinfo{volume}{10}},
  \bibinfo{pages}{303} (\bibinfo{year}{1999}{\natexlab{a}}),
  \eprint{astro-ph/9807045}.

\bibitem[{\citenamefont{Hayashida
  et~al.}(1999{\natexlab{b}})}]{Hayashida:1999ab}
\bibinfo{author}{\bibfnamefont{N.}~\bibnamefont{Hayashida}}
  \bibnamefont{et~al.} (\bibinfo{collaboration}{AGASA})
  (\bibinfo{year}{1999}{\natexlab{b}}), \bibinfo{note}{proc. 26th International
  Cosmic Ray Conference (ICRC 99), Salt Lake City, USA},
  \eprint{astro-ph/9906056}.

\bibitem[{\citenamefont{Bellido et~al.}(2001)\citenamefont{Bellido, Clay,
  Dawson, and Johnston-Hollitt}}]{Bellido:2000tr}
\bibinfo{author}{\bibfnamefont{J.~A.} \bibnamefont{Bellido}},
  \bibinfo{author}{\bibfnamefont{R.~W.} \bibnamefont{Clay}},
  \bibinfo{author}{\bibfnamefont{B.~R.} \bibnamefont{Dawson}},
  \bibnamefont{and}
  \bibinfo{author}{\bibfnamefont{M.}~\bibnamefont{Johnston-Hollitt}},
  \bibinfo{journal}{Astropart. Phys.} \textbf{\bibinfo{volume}{15}},
  \bibinfo{pages}{167} (\bibinfo{year}{2001}), \eprint{astro-ph/0009039}.

\bibitem[{\citenamefont{Santos}(2007)}]{Santos:2007na}
\bibinfo{author}{\bibfnamefont{E.~M.} \bibnamefont{Santos}}
  (\bibinfo{collaboration}{Pierre Auger}) (\bibinfo{year}{2007}),
  \bibinfo{note}{proc. 30th International Cosmic Ray Conference (ICRC 2007),
  Merida, Mexico}, \eprint{0706.2669}.

\bibitem[{\citenamefont{Stanev et~al.}(1995)\citenamefont{Stanev, Biermann,
  Lloyd-Evans, Rachen, and Watson}}]{Stanev:1995my}
\bibinfo{author}{\bibfnamefont{T.}~\bibnamefont{Stanev}},
  \bibinfo{author}{\bibfnamefont{P.~L.} \bibnamefont{Biermann}},
  \bibinfo{author}{\bibfnamefont{J.}~\bibnamefont{Lloyd-Evans}},
  \bibinfo{author}{\bibfnamefont{J.~P.} \bibnamefont{Rachen}},
  \bibnamefont{and} \bibinfo{author}{\bibfnamefont{A.~A.}
  \bibnamefont{Watson}}, \bibinfo{journal}{Phys. Rev. Lett.}
  \textbf{\bibinfo{volume}{75}}, \bibinfo{pages}{3056} (\bibinfo{year}{1995}),
  \eprint{astro-ph/9505093}.

\bibitem[{\citenamefont{Hayashida et~al.}(1996)}]{Hayashida:1996bc}
\bibinfo{author}{\bibfnamefont{N.}~\bibnamefont{Hayashida}}
  \bibnamefont{et~al.}, \bibinfo{journal}{Phys. Rev. Lett.}
  \textbf{\bibinfo{volume}{77}}, \bibinfo{pages}{1000} (\bibinfo{year}{1996}).

\bibitem[{\citenamefont{Kewley et~al.}(1996)\citenamefont{Kewley, Clay, and
  Dawson}}]{Kewley:1996zt}
\bibinfo{author}{\bibfnamefont{L.~J.} \bibnamefont{Kewley}},
  \bibinfo{author}{\bibfnamefont{R.~W.} \bibnamefont{Clay}}, \bibnamefont{and}
  \bibinfo{author}{\bibfnamefont{B.~R.} \bibnamefont{Dawson}},
  \bibinfo{journal}{Astropart. Phys.} \textbf{\bibinfo{volume}{5}},
  \bibinfo{pages}{69} (\bibinfo{year}{1996}).

\bibitem[{\citenamefont{Bird et~al.}(1998)}]{Bird:1998nu}
\bibinfo{author}{\bibfnamefont{D.~J.} \bibnamefont{Bird}} \bibnamefont{et~al.}
  (\bibinfo{collaboration}{HiRes}) (\bibinfo{year}{1998}),
  \eprint{astro-ph/9806096}.

\bibitem[{\citenamefont{Kashti and Waxman}(2008)}]{Kashti:2008bw}
\bibinfo{author}{\bibfnamefont{T.}~\bibnamefont{Kashti}} \bibnamefont{and}
  \bibinfo{author}{\bibfnamefont{E.}~\bibnamefont{Waxman}},
  \bibinfo{journal}{JCAP} \textbf{\bibinfo{volume}{0805}}, \bibinfo{pages}{006}
  (\bibinfo{year}{2008}), \eprint{0801.4516}.

\bibitem[{\citenamefont{Koers and Tinyakov}(2009{\natexlab{a}})}]{Koers:2008ba}
\bibinfo{author}{\bibfnamefont{H.~B.~J.} \bibnamefont{Koers}} \bibnamefont{and}
  \bibinfo{author}{\bibfnamefont{P.}~\bibnamefont{Tinyakov}},
  \bibinfo{journal}{JCAP} \textbf{\bibinfo{volume}{0904}}, \bibinfo{pages}{003}
  (\bibinfo{year}{2009}{\natexlab{a}}), \eprint{0812.0860}.

\bibitem[{\citenamefont{Abu-Zayyad et~al.}(1999)}]{HiResStatus1999}
\bibinfo{author}{\bibfnamefont{T.}~\bibnamefont{Abu-Zayyad}}
  \bibnamefont{et~al.} (\bibinfo{collaboration}{HiRes}) (\bibinfo{year}{1999}),
  \bibinfo{note}{in: Proc. 26th ICRC, 4, 349 (1999) [check reference with GT:
  cannot find in spires; ads gives volume 5]}.

\bibitem[{\citenamefont{Boyer et~al.}(2002)\citenamefont{Boyer, Knapp, Mannel,
  and Seman}}]{Boyer:2002zz}
\bibinfo{author}{\bibfnamefont{J.~H.} \bibnamefont{Boyer}},
  \bibinfo{author}{\bibfnamefont{B.~C.} \bibnamefont{Knapp}},
  \bibinfo{author}{\bibfnamefont{E.~J.} \bibnamefont{Mannel}},
  \bibnamefont{and} \bibinfo{author}{\bibfnamefont{M.}~\bibnamefont{Seman}},
  \bibinfo{journal}{Nucl. Instrum. Meth.} \textbf{\bibinfo{volume}{A482}},
  \bibinfo{pages}{457} (\bibinfo{year}{2002}).

\bibitem[{\citenamefont{Hanlon}(2008)}]{Hanlon:2008}
\bibinfo{author}{\bibfnamefont{W.}~\bibnamefont{Hanlon}}, Ph.D. thesis,
  \bibinfo{school}{University of Utah} (\bibinfo{year}{2008}),
  \bibinfo{note}{http://www.cosmic-ray.org/thesis/hanlon.html}.

\bibitem[{\citenamefont{Abbasi et~al.}(2008{\natexlab{b}})}]{Abbasi:2008md}
\bibinfo{author}{\bibfnamefont{R.~U.} \bibnamefont{Abbasi}}
  \bibnamefont{et~al.}, \bibinfo{journal}{Astropart. Phys.}
  \textbf{\bibinfo{volume}{30}}, \bibinfo{pages}{175}
  (\bibinfo{year}{2008}{\natexlab{b}}), \eprint{0804.0382}.

\bibitem[{\citenamefont{Abbasi et~al.}(2009)}]{Abbasi:2009nf}
\bibinfo{author}{\bibfnamefont{R.~U.} \bibnamefont{Abbasi}}
  \bibnamefont{et~al.} (\bibinfo{year}{2009}), \eprint{arXiv 0910.4184}.

\bibitem[{\citenamefont{Huchra et~al.}(2009)\citenamefont{Huchra, Macri,
  Jarrett, Martimbeau, Masters, Crook, Cutri, Schneider, and Skutskie}}]{2MRS}
\bibinfo{author}{\bibfnamefont{J.}~\bibnamefont{Huchra}},
  \bibinfo{author}{\bibfnamefont{L.}~\bibnamefont{Macri}},
  \bibinfo{author}{\bibfnamefont{T.}~\bibnamefont{Jarrett}},
  \bibinfo{author}{\bibfnamefont{N.}~\bibnamefont{Martimbeau}},
  \bibinfo{author}{\bibfnamefont{K.}~\bibnamefont{Masters}},
  \bibinfo{author}{\bibfnamefont{A.}~\bibnamefont{Crook}},
  \bibinfo{author}{\bibfnamefont{R.}~\bibnamefont{Cutri}},
  \bibinfo{author}{\bibfnamefont{S.}~\bibnamefont{Schneider}},
  \bibnamefont{and} \bibinfo{author}{\bibfnamefont{M.}~\bibnamefont{Skutskie}}
  (\bibinfo{year}{2009}), \bibinfo{note}{in preparation}.

\bibitem[{\citenamefont{{Koers} and {Tinyakov}}(2009{\natexlab{b}})}]{Koers:2009pd}
\bibinfo{author}{\bibfnamefont{H.~B.~J.} \bibnamefont{{Koers}}}
  \bibnamefont{and}
  \bibinfo{author}{\bibfnamefont{P.}~\bibnamefont{{Tinyakov}}},
  \bibinfo{journal}{Mon. Not. Roy. Astron. Soc.}
  \textbf{\bibinfo{volume}{399}}, \bibinfo{pages}{1005} (\bibinfo{year}{2009}{\natexlab{b}}),
  \eprint{0907.0121}.

\bibitem[{\citenamefont{Koers and Tinyakov}(2008)}]{Koers:2008hv}
\bibinfo{author}{\bibfnamefont{H.~B.~J.} \bibnamefont{Koers}} \bibnamefont{and}
  \bibinfo{author}{\bibfnamefont{P.}~\bibnamefont{Tinyakov}},
  \bibinfo{journal}{Phys. Rev.} \textbf{\bibinfo{volume}{D78}},
  \bibinfo{pages}{083009} (\bibinfo{year}{2008}), \eprint{0802.2403}.

\bibitem[{\citenamefont{Abbasi et~al.}(2007)}]{Abbasi:2006mb}
\bibinfo{author}{\bibfnamefont{R.}~\bibnamefont{Abbasi}} \bibnamefont{et~al.}
  (\bibinfo{collaboration}{HiRes}), \bibinfo{journal}{Astropart. Phys.}
  \textbf{\bibinfo{volume}{27}}, \bibinfo{pages}{370} (\bibinfo{year}{2007}),
  \eprint{astro-ph/0607094}.

\bibitem[{\citenamefont{Bergman}(2007)}]{Bergman:2006vt}
\bibinfo{author}{\bibfnamefont{D.~R.} \bibnamefont{Bergman}}
  (\bibinfo{collaboration}{HiRes}), \bibinfo{journal}{Nucl. Phys. B Proc.
  Suppl.} \textbf{\bibinfo{volume}{165}}, \bibinfo{pages}{19}
  (\bibinfo{year}{2007}), \eprint{astro-ph/0609453}.

\bibitem[{\citenamefont{Bird et~al.}(1993)}]{Bird:1993yi}
\bibinfo{author}{\bibfnamefont{D.~J.} \bibnamefont{Bird}} \bibnamefont{et~al.}
  (\bibinfo{collaboration}{HiRes}), \bibinfo{journal}{Phys. Rev. Lett.}
  \textbf{\bibinfo{volume}{71}}, \bibinfo{pages}{3401} (\bibinfo{year}{1993}).

\bibitem[{\citenamefont{Abu-Zayyad et~al.}(2000)}]{AbuZayyad:2000zz}
\bibinfo{author}{\bibfnamefont{T.}~\bibnamefont{Abu-Zayyad}}
  \bibnamefont{et~al.}, \bibinfo{journal}{Phys. Rev. Lett.}
  \textbf{\bibinfo{volume}{84}}, \bibinfo{pages}{4276} (\bibinfo{year}{2000}).

\bibitem[{\citenamefont{Abu-Zayyad et~al.}(2001)}]{AbuZayyad:2000ay}
\bibinfo{author}{\bibfnamefont{T.}~\bibnamefont{Abu-Zayyad}}
  \bibnamefont{et~al.} (\bibinfo{collaboration}{HiRes-MIA}),
  \bibinfo{journal}{Astrophys. J.} \textbf{\bibinfo{volume}{557}},
  \bibinfo{pages}{686} (\bibinfo{year}{2001}), \eprint{astro-ph/0010652}.

\bibitem[{\citenamefont{Abbasi et~al.}(2005)}]{Abbasi:2004nz}
\bibinfo{author}{\bibfnamefont{R.~U.} \bibnamefont{Abbasi}}
  \bibnamefont{et~al.} (\bibinfo{collaboration}{HiRes}),
  \bibinfo{journal}{Astrophys. J.} \textbf{\bibinfo{volume}{622}},
  \bibinfo{pages}{910} (\bibinfo{year}{2005}), \eprint{astro-ph/0407622}.

\bibitem[{\citenamefont{Heck et~al.}(1998)\citenamefont{Heck, Schatz, Thouw,
  Knapp, and Capdevielle}}]{Heck:1998vt}
\bibinfo{author}{\bibfnamefont{D.}~\bibnamefont{Heck}},
  \bibinfo{author}{\bibfnamefont{G.}~\bibnamefont{Schatz}},
  \bibinfo{author}{\bibfnamefont{T.}~\bibnamefont{Thouw}},
  \bibinfo{author}{\bibfnamefont{J.}~\bibnamefont{Knapp}}, \bibnamefont{and}
  \bibinfo{author}{\bibfnamefont{J.~N.} \bibnamefont{Capdevielle}}
  (\bibinfo{year}{1998}), \bibinfo{note}{tech. Rep. FZKA 6019,
  Forschungszentrum Karlsruhe}.

\bibitem[{\citenamefont{Kalmykov et~al.}(1997)\citenamefont{Kalmykov,
  Ostapchenko, and Pavlov}}]{Kalmykov:1997te}
\bibinfo{author}{\bibfnamefont{N.~N.} \bibnamefont{Kalmykov}},
  \bibinfo{author}{\bibfnamefont{S.~S.} \bibnamefont{Ostapchenko}},
  \bibnamefont{and} \bibinfo{author}{\bibfnamefont{A.~I.}
  \bibnamefont{Pavlov}}, \bibinfo{journal}{Nucl. Phys. B Proc. Suppl.}
  \textbf{\bibinfo{volume}{52B}}, \bibinfo{pages}{17} (\bibinfo{year}{1997}).

\bibitem[{\citenamefont{{V{\'e}ron-Cetty} and
  {V{\'e}ron}}(2006)}]{2006A&A...455..773V}
\bibinfo{author}{\bibfnamefont{M.-P.} \bibnamefont{{V{\'e}ron-Cetty}}}
  \bibnamefont{and}
  \bibinfo{author}{\bibfnamefont{P.}~\bibnamefont{{V{\'e}ron}}},
  \bibinfo{journal}{Astron. Astrophys.} \textbf{\bibinfo{volume}{455}},
  \bibinfo{pages}{773} (\bibinfo{year}{2006}).

\end{thebibliography}

\end{document}